\documentclass[11pt]{article} 

\usepackage[utf8]{inputenc} 

\usepackage{geometry} 
\usepackage{graphicx} 

\usepackage{booktabs} 
\usepackage{array} 
\usepackage{paralist} 
\usepackage{verbatim} 
\usepackage{subfig} 

\usepackage{fancyhdr} 
\usepackage{sectsty}
\allsectionsfont{\sffamily\mdseries\upshape} 

\usepackage[nottoc,notlof,notlot]{tocbibind} 
\usepackage[titles,subfigure]{tocloft} 

\usepackage{upgreek}
\usepackage{hyperref}

\title{Transport of Intensity Equation Microscopy for Dynamic Microtubules.}
\author{Q. Tyrell Davis\textsuperscript{1,2,*}}

\begin{document}
\maketitle
\begin{center}
\noindent
\textbf{1} Centre for Gene Regulation and Expression, School of Life Sciences, University of Dundee, Dundee, UK
\\
\textbf{2} PHOQUS (https://www.phoqus.eu/) \\
\textbf{*} q.t.z.davis@dundee.ac.uk,
\end{center}

\section*{Abstract} 

Microtubules (MTs) are filamentous protein polymers roughly 25 nm in diameter. Ubiquitous in eukaryotes, MTs are well known for their structural role but also act as actuators, sensors, and, in association with other proteins, checkpoint regulators. The thin diameter and transparency of microtubules classifies them as sub-resolution phase objects, with concomitant imaging challenges. Label-free methods for imaging microtubules are preferred when long exposure times would lead to phototoxicity in fluorescence, or for retaining more native structure and activity. 

This method approaches quantitative phase imaging of MTs as an inverse problem based on the Transport of Intensity Equation. In a co-registered comparison of MT signal-to-background-noise ratio, TIE Microscopy of MTs shows an improvement of more than three times that of video-enhanced bright field imaging.

This method avoids the anisotropy caused by prisms used in differential interference contrast and takes only two defocused images as input. Unlike other label-free techniques for imaging microtubules, in TIE microscopy background removal is a natural consequence of taking the difference of two defocused images, so the need to frequently update a background image is eliminated. 


\section*{Introduction}

Microtubules (MTs) play essential roles in biological processes including structure, transport, motility, and cell division; MT defects are associated with a broad range of fundamental pathologies \cite{bailey2013}.

MTs form the framework for many classical and current {\itshape in vitro} experiments, including biophysical experiments involving force measurement with optical tweezers \cite{block1990,bailey2013}, for which extended live-viewing requirements make label-free MT imaging techniques a requirement.

A variety of label-free methods for imaging MTs have been demonstrated previously: dark-field \cite{horio1986}, video enhanced image processing with \cite{soboeiro1988} or without \cite{medina2010} differential interference contrast,  quantitative polarisation \cite{oldenbourg1998}, reflected methods based on interference \cite{amos2011,andrecka2016}, and phase contrast \cite{kandel2017} 
have all been demonstrated for {\itshape in vitro} microtubule experiments. Despite the availability of label-free methods, fluorescence microscopy continues to dominate {\itshape in vitro} microtubule imaging thanks to high contrast and specificity. However, fluorescence imaging intrinsically adds two non-negligible perturbations: phototoxicity and photobleaching, which can damage microtubules \cite{guo2006}. Even in the absence of excitation illumination, labeling can alter MT activity and dynamic interactions \cite{kandel2017}. 
Consequently, experiments that require extended timelapse or constant live viewing make especially poor candidates for fluorescence microscopy. 

Label-free imaging with video-enhanced DIC (VE-DIC) is widely used in conjunction with optical tweezers \cite{bailey2013}, and more recently video-enhanced bright field imaging was demonstrated using the same image processing \cite{candia2013}. The latter avoids the distortion of optical traps caused by DIC prisms, which can cause anisotropies in trap stiffness of about 30\% as described in  \cite{lang2002}. An alternative solution to trap anisotropy is to employ a 4f relay to physically remove the DIC prism from the trapping beam path as in \cite{deng2017}. In practice, maintaining sufficient contrast in VE-DIC to image single MTs requires frequent prism re-alignment.
 
Bright field microscopy, combined with similar image processing to that employed in VE-DIC, avoids VE-DIC anisotropies at the cost of lower contrast \cite{medina2010,candia2013}. As pure phase objects, contrast at best focus is negligible for MTs, and in Video-Enhanced Bright Field (VE-BF) MTs are necessarily defocused for imaging as noted in \cite{candia2013}. This defocus-based contrast inverts from dark to light as the phase object passes from above to below the focal plane. Contrast, {\itshape i.e.} the intensity of a phase object image in relation to background noise, propagates in a predictable way described by the Transport of Intensity Equation (TIE, Eq. \ref{eq:TIE}).

TIE microscopy of MTs maintains the unperturbed optical path of VE-BF, does away with a requirement for dedicated background acquisition and subtraction, and improves contrast over VE-BF to be comparable to VE-DIC. This is achieved by solving the inverse problem of light intensity due to defocus to compute phase map images of microtubules.

TIE microscopy requires two or more defocused images as inputs. 
Defocused images for TIEM can be obtained by physically moving a sample stage \cite{barty1998}, refocusing an optical element (such as a tuneable lens \cite{zuo2013}), or one of several single-shot multifocus options \cite{abrahamsson2012,waller2010}. In this demonstration I use a precision piezo stage to move the sample to several different defocus positions..
 
\begin{figure}[!h]
\includegraphics[keepaspectratio,scale=0.42]{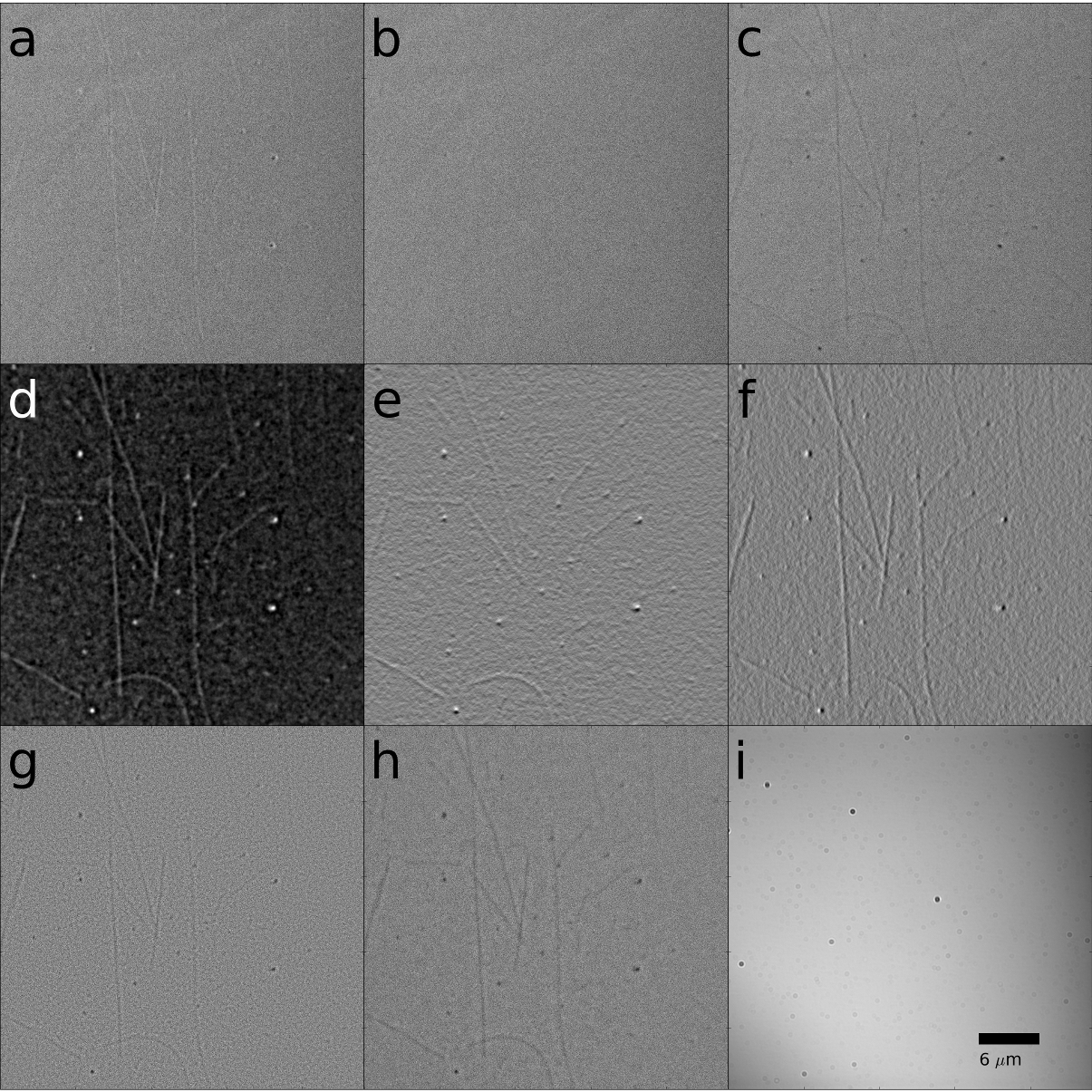}
\caption{{\bf TIEM of Microtubules and Comparison to Other Imaging Modes} \\
{\bf (a)} Video-enhanced bright field micrograph of microtubules adhered to a coverslip (average of 42 frames with background subtracted), defocused 225 nm below the sample plane ($I_b$). {\bf (b)} VE-BF at the sample plane ($I_0$). {\bf (c)} VE-BF  225 nm above the sample plane ($I_a$). {\bf (d)} TIE phase contrast micrograph of microtubules. {\bf (e)} The gradient of {\bf d} mimicking DIC contrast with a y-axis shear. {\bf (f)} Gradient of {\bf d} mimicking DIC contrast with a x-axis shear. Contrast in {\bf e} and {\bf f} is essentially nonexistent for MTs parallel to the shear axis.
{\bf (g)} The Laplacian ($\nabla^2[\phi(x,y)]$) of {\bf d}. 
{\bf (h)} The intensity difference image ($I_a-I_b$). For micrographs of sufficiently thin phase objects, the first term in Eq. \ref{eq:productrule} goes to 0 and {\bf h} and {\bf g} are mathematically equivalent. 
{\bf (i)} `Through the eyepiece' (no background subtraction) view of the sample plane.}
\label{fig:vissummary}
\end{figure}
 
\section*{Materials and Methods}

\subsection*{Transport of intensity in the case of microtubules}

The transport of intensity equation (TIE) describes the change in intensity of a wavefront that passes through a thin phase object at different planes of focus. This was derived by Teague in 1983 \cite{teague1983}, and later applied to visible light microscopy \cite{barty1998}.

The TIE describes the first derivative of intensity with respect to focus $z$ as: 

\begin{equation}
k_0 \frac{\partial I(x,y,z)} {\partial z}  = - \nabla [I_{z=0}(x,y,z) \nabla (\phi_{z=0}(x,y,z)) ] 
\label{eq:TIE}
\end{equation}

Where $I_{z=0}(x,y,z)$ is the intensity at the focal plane, $\phi(x,y,z)$ is the phase, and $k_0 = \frac{2\pi}{\lambda}$ is the wavenumber. The first derivative with respect to focus is measured empirically as the difference of two images, $I_a$ and $I_b$, focused above and below the sample plane, respectively. 

We can follow Teague \cite{teague1983} in introducing an additional function $\psi(x,y,z)$ defined as $ \nabla(\psi(x,z,y)) := I_{z=0}(x,y,z)\nabla(\phi_{z=0}(x,y,z)) $, so that the right hand side of Eq \ref{eq:TIE} becomes a Poisson's equation.

\begin{equation}
k_0\frac{I_a-I_b}{\Delta z} = - \nabla^2 ( \psi(x,y,z))
\end{equation}

This simplification allows us to approach the problem of TIE phase retrieval with our choice of Poisson solver. A simple and computationally fast approach takes advantage of the Fourier transform pair for the Laplacian operator as in \cite{gureyev1997}. After taking the Fourier transform, rearranging to solve for $\Psi$ yields: 
\begin{equation}
\Psi(k_x,k_y) =- \frac{ k_0\mathcal{F}[\frac{\partial I}{\partial z}]}{4\pi^2(k_x^2+k_y^2)}
\label{eq:FourierPair}
\end{equation}

As spatial frequencies approach the DC bias term, we run the risk of dividing by zero due to the presence of $k_x^2 +k_y^2$ in the denominator. Low spatial frequency inaccuracies in the estimate of $\frac{\partial I}{\partial z}$ will be disproportionately amplified as a result, which can be seen as a fog-like artifact in some TIE images. To prevent division by zero and decrease sensitivity to noise, we can add an offset in the Fourier domain. To restrict the effect of our offset to low spatial frequencies, we add this offset in the form of a 2D Hanning window modified to center the window at DC frequency. Adding the Hanning window attenuates the low-pass filter effects in the Fourier domain as well as avoiding division by zero. This is especially well suited to our objective of imaging sub-resolution phase objects such as MTs, because these are represented at high frequencies at the limits of the imaging system's capabilities. 


For other imaging situations, such as cells or microspheres, the scale of the Hanning function can be adjusted to retain more low spatial frequencies as required. The Hanning function centered at zero frequency is described as:

\begin{equation}
k_r = \sqrt{k_x^2 + k_y^2}
\end{equation}

\begin{equation}
f_{Hann}(k_r)|_{k_r \leq \omega_0} = \frac{\alpha}{2} (1+cos( \frac{\pi k_r}{\omega_0} )) 
\end{equation}

\begin{equation}
f_{Hann}(k_r)|_{k_r \geq \omega_0} = 0
\label{eq:hanning}
\end{equation}

Where $\omega_0$ is the spatial frequency at which the Hanning function goes to 0 and $\alpha$ is a scaling factor.

\begin{figure}[!h]
\includegraphics[keepaspectratio,scale=.225]{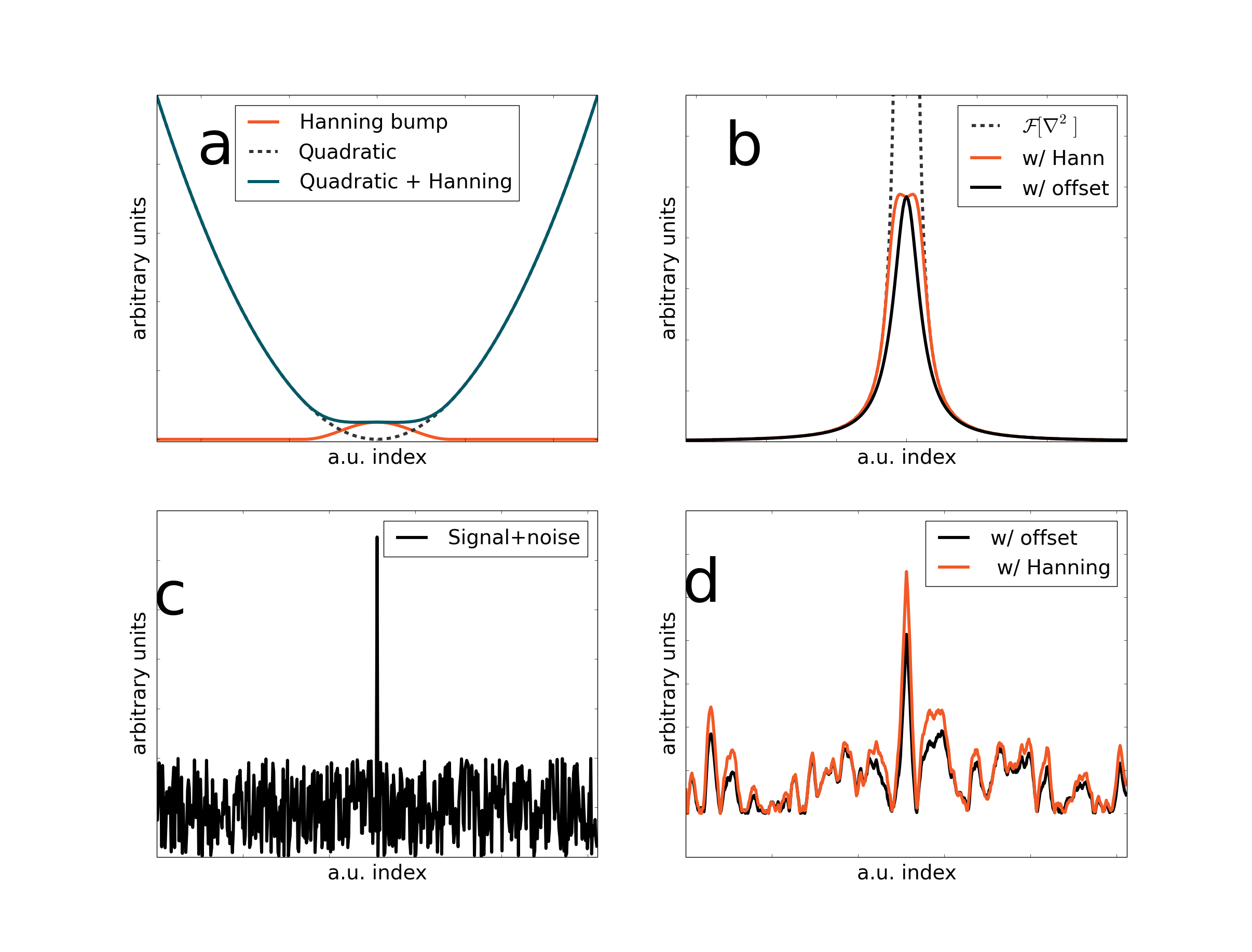}
\caption{{\bf A Hanning Bump Stabilizes the TIE in the Fourier domain.} This figure demonstrates how solving the TIE amplifies large objects and attenuates small objects and noise. In the Fourier domain large image objects occur near 0 spatial frequency, smaller objects are represented by higher spatial frequencies. In these cross sections 0 spatial frequency has been shifted to the center.
{\bf (a)} The quadratic in the Fourier domain Laplacian illustrates the instability of the TIE: the result will blow up at low spatial frequencies as the denominator approaches 0. Adding a global offset prevents the denominator from reaching 0, but a Hanning bump can be used to smoothly and specifically adjust the TIE, resulting in less deviation from the analytical solution. {\bf (b)} Demonstrates the effect of adding a Hanning bump or a global offset to prevent division by 0. {\bf (c)} Simulated signal with added noise. The spike in the middle, a thin Gaussian function, has an amplitude 8 times the standard deviation of the background noise. {\bf (d)} Simulated signal plus noise after convolution with {\bf b}, illustrating the effect of using a Hanning bump instead of a global offset. An interactive version of this figure is available in Figure S3.}

\label{fig:VEBFTIELap}
\end{figure}

By taking the inverse Fourier transform of the right hand side of Eq \ref{eq:FourierPair} we are left with the auxiliary function $\psi$ we introduced earlier. We can solve for $\phi$ by integrating as Teague suggests \cite{teague1983}, or reuse our Fourier domain Poisson solver, taking the gradient of the quotient of $\psi$ and $I_{z=0}$ to generate a new Poisson's equation, this one yielding $\phi$ when solved as above \cite{gorthi2012}. 

Once we have a phase map, $\phi(x,y)$ of our sample plane, we can process the complex wavefront  {\itshape in silico} yielding diverse imaging modes (Fig. \ref{fig:vissummary}). DIC imaging is sensitive to the gradient of phase, so we can simulate DIC microscopy with a chosen shear axis by taking the gradient of our phase image \cite{lue2007}.
Applying the Laplacian operator to the phase image, we can highlight small features with improved isotropic contrast. Laplace Phase Microscopy (LPM) has been shown to be advantageous for label-free study of the fine details of intracellular organelle transport \cite{wang2011}. When imaging MTs, we can use the intensity difference of two symmetrically defocused images as a close estimate of the Laplacian of phase micrograph thanks to a particular simplification of the TIE for thin phase objects.

A visual inspection of a Laplacian micrograph and the defocused image $I_a - I_b$ (Fig \ref{fig:vissummary}) reveals a striking similarity. We can apply the product rule for gradients to Eq \ref{eq:TIE} to describe the TIE in two terms (dropping the coordinate dependencies here for simplicity): 

\begin{equation}
k_0 \frac{\partial I} {\partial z}  = -[ \nabla (I_{z=0})\nabla(\phi_{z=0}) +  I_{z=0}\nabla^2 (\phi_{z=0})] 
\label{eq:productrule}
\end{equation}

For a sufficiently thin pure phase object, contrast is minimal when the sample is at best focus (as noted by Frits Zernike in 1955 \cite{zernike}, where he summarizes earlier work \cite{abbe} by Ernst Abbe). Zernike noted that a microscopist working with unstained cells is likely to continuously and unconsciously focus in fine increments about the sample plane, building a mental picture of the phase object. This is similar to what we aim to accomplish here computationally. 

At best focus a thin phase object visualized by a microscope creates a uniform, featureless image, and the corresponding gradient is zero everywhere, except for contributions from the imaging system and due to noise. Non-uniform illumination due to the imaging system is contained in both defocused images, and is removed by taking their difference. Therefore we can eliminate the first term in Eq \ref{eq:productrule} and replace $I_0(x,y)$ with its mean, a constant. We then use our Fourier Poisson solver to recover $\phi$, and the defocus image $\frac{\partial I}{\partial z}$ can act as a low-computation substitute for the Laplacian phase micrograph calculated from $\phi(x,y)$ in Fig \ref{fig:vissummary}, because for sufficiently thin phase objects the difference of two defocused microscopical images is proportional to the Laplacian of the optical path length of the object. This relationship between the difference of defocused images and the Laplacian of phase has been noted before as the premise for defocusing microscopy, and demonstrated as a means for studying membrane fluctuations \cite{agero2004}.

As described above, the first derivative of intensity can be estimated by the difference of two images at different levels of defocus. More images can be used to improve the estimate of $\frac{\partial I}{\partial z}$, as has been described previously \cite{soto2007,waller2010} and the estimate of $\frac{\partial I}{\partial z}$ can be improved by considering the noise level and magnitude of intensity transport for a given imaging context \cite{carranza2014,jingshan2014}. For the purposes of this paper, I used two images defocused by distances of 200 to 500 nm to estimate $\frac{\partial I}{\partial z}$ for computing TIE phase maps.   

\subsubsection*{Microscope Setup}

\begin{figure}[!h]

\includegraphics[keepaspectratio,scale=0.35]{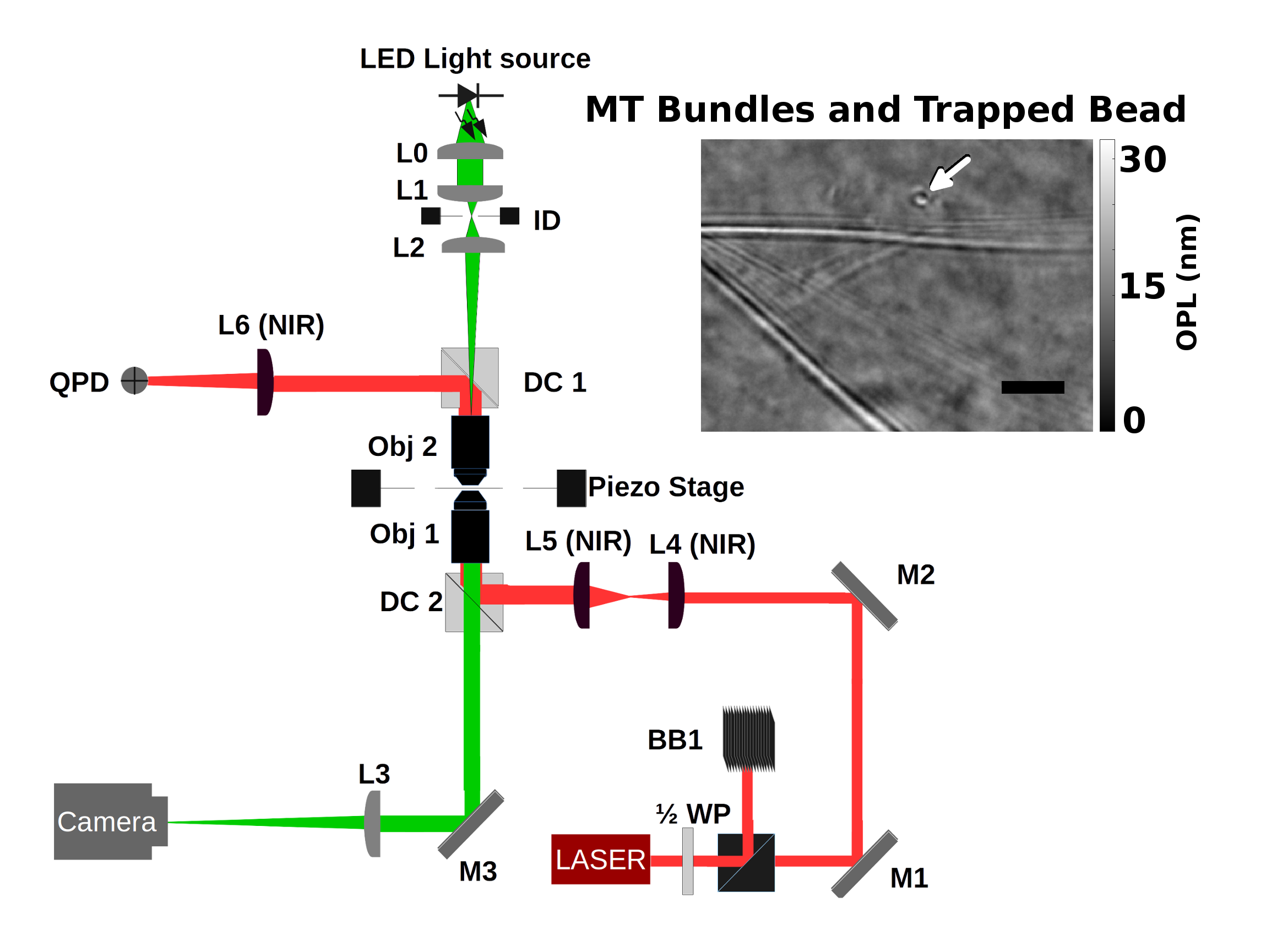}
\caption{{\bf Cartoon Representation of Optical Tweezers Microscope}
The microscope used to demonstrate TIE microscopy of microtubules is part of an optical tweezers system. The trapping arm, not used in this paper, is shown in red and the imaging arm is shown in green. {\bf Inset:} Microsphere trapped by optical tweezers and held in the vicinity of microtubule bundles, imaged with TIE microscopy. An arrow indicates the trapped microsphere, scale bar 5 $\upmu$m.
}
\label{fig:TIEMCartoon}
\end{figure}

A cartoon representation of the microscope used in these experiments, part of a custom optical tweezers system, is shown in Figure \ref{fig:TIEMCartoon}. Three lenses (L0, L1, and L2) are used to conjugate the image of the LED chip to an iris diaphragm  (ID 1), which is in turn conjugated to the back focal plane of the illumination objective, providing K\"ohler illumination. The illumination objective is a high NA oil immersion objective (Nikon E Plan 100X, NA = 1.25) for purposes of collecting the trapping beam of the optical tweezers system, but the effective illumination NA is stopped down using the iris diaphragm ID. Limiting the illumination increases depth-of-field and illumination coherence, improving the resulting TIE phase map. A piezo stage moves the sample plane to provide defocus (BIO3.100, Piezoconcept, Lyon, France). The sample plane is imaged onto a camera (Mako G-125, www.alliedvision.com) by the imaging objective (Nikon Plan Fluor 100X, NA = 1.3) and lens L3 (f = 200 mm).

\subsubsection*{SNR Measurements}
Signal to noise ratios were measured according to the definition presented in \cite{bormuth2007}. Using a MATLAB script, I measured the peak-to-peak value of the average of 21 MT cross-section profile plots and divided the result by the standard deviation of a rectangular region of nearby background, a 20 by 20 pixels area free from microtubules and blobs on the coverslip. To avoid attributing SNR differences due to variation across images to the method used, I measured co-registered sections of MTs at the same timepoint. The reported frames are the total frames used to create each image, excepting the additional requirement for VE-BF of a background image (an average of many frames acquired at 3 microns negative defocus). 

\subsubsection*{Preparation of Stable Microtubule Seeds}

Stabilized microtubule seeds were prepared by incubating a 3.8 $\upmu$M porcine tubulin mixture with 1 mM guanylyl-($\upalpha,\upbeta$)-methylenediphosphate (GMPCPP, a non-hydrolysable GTP analogue) at 37$^\circ$C for one hour in MRB80 buffer (80 mM Pipes, 4 mM MgCl$_2$, and 1 mM EGTA, pH 6.8). Before incubation, the mixture was left on ice for 5 minutes to ensure tubulin depolymerization. I then spun the mix on an airfuge (www.beckmancoulter.com, 347855) at 30 psi for 5 minutes to remove un-polymerized tubulin, resuspending the resulting transparent pellet in 200 $\upmu$l MRB80. Microtubule seeds were stable at room temperature for a week or longer, and were used within 4 days for experiments described in this document.

\subsubsection*{Unlabeled/Labeled Tubulin Dynamic MT Timelapse Experiments}

Zeiss  \#1 1/2 coverslips (www.micro-shop.zeiss.com, 474030-9000-000 and 474030-9020-000) were plasma-cleaned in a ceramic holder for two cycles of 30 s at 10 mA using a Turbo Carbon Coater and Auxiliary Power Unit (www.agarscientific.com, AGB7230 and AGB7252).  The cleaned coverslips were then silanized by placing in a solution of 125 $\upmu$l Dichlorodimethylsilane (www.sigmaaldrich.com, 440272) in ~250 ml Trichlorethylene  (www.sigmaaldrich.com, 251402)  for 1 hour. The silanization solution was removed and the coverslips were placed in fresh methanol (www.VWR.com, 20847.307) for three subsequent rounds of sonication of 5, 15, and 30 minutes in a sonicator bath.

To construct flow chambers, I cut channels out of Parafilm (www.parafilm.com, PM-996) and sandwiched them between 18x18 and 22x22 mm cleaned and silanized coverslips, sealing the flow chambers by melting the Parafilm on a 100$^\circ$C heating block. The glass/Paraflim/glass stack  transitions from cloudy to clear in appearance when sealed. Each channel holds less than 10 $\upmu$l volume. 

Channels were filled with 40 $\upmu$l of MRB80, then incubated with 10 $\upmu$l of 200X diluted anti-$\upbeta$ antibodies (www.sigmaaldrich.com, T8578). After a wash with 40 $\upmu$l of MRB80, 10 $\upmu$l 1\% pluronic F127 was flowed into the channel. Another 40 $\upmu$l MRB80 wash was followed by an incubation with 10 $\upmu$l of stable microtubule seeds, followed by a final wash and the introduction of 20 $\upmu$l of dynamic MT reaction mixture. All incubations were 5 minutes at room temperature.

Dynamic MT reaction mix was made by combining 15 $\upmu$M porcine tubulin (or 14.1 $\upmu$M porcine tubulin with 0.9 $\upmu$M Rhodamine-labeled tubulin  from www.cytoskeleton.com, TL590M), with 50 mM KCl, 0.5 mg/ml $\upkappa$-casein, 1 mM GTP, 20 mM glucose, and oxygen scavenger mix  (400 $\upmu$g/mL glucose oxidase, 200 $\upmu$g/mL catalase, and 4 mM DTT). The final volume for the reaction mix was 20 $\upmu$l, the remainder of which was MRB80 buffer. 

\subsubsection*{Analysis of Dynamic Timelapse Images}
Dynamic MT timelapses were analyzed in ImageJ  (https://imagej.nih.gov/ij/)\cite{rasband2016}, catastrophes were counted in blinded timelapse images by volunteers otherwise uninvolved in the study, and dynamic MT traces in Figure \ref{fig:dynMTs} were measured using ImageJ's line segment measure tool. I used the R statistical package to perform a Poisson statistical test to compare labeled and unlabeled catastrophe rates \cite{R}. Timelapse imaging was conducted at room temperature.

\section*{Results and Discussion}

\subsection*{Rhodamine-Labeled Dynamic Microtubules Have a Lower Catastrophe Frequency}

\begin{figure}[!h]
\includegraphics[keepaspectratio,scale=0.5]{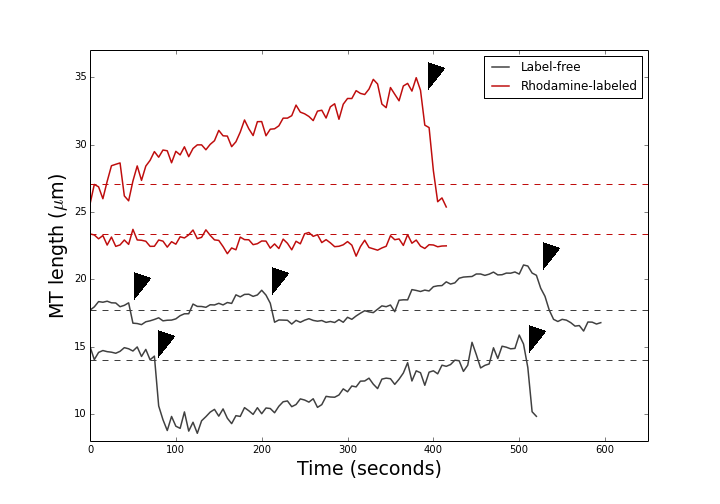}
\caption{{\bf Dynamic MT tracking with TIE phase imaging} Dynamic lengths of individual microtubules, demonstrating periods of polymerization, catastrophe, and pausing. Traces are manually offset to make it easy to distinguish the traces for each individual MT.  Black arrows indicate catastrophes.} 
\label{fig:dynMTs}
\end{figure}

To demonstrate the utility of TIE Microscopy for imaging dynamic microtubules I performed time lapse imaging of coverslip-attached dynamic microtubules. I chose catastrophe frequency as a metric to determine if MT dynamics differ for unlabeled vs labeled tubulin, as I expect blinded analysis of catastrophes to be more objective then measurements of growth and depolymerization.
Figure \ref{fig:dynMTs} 
shows typical timelapse length measurements of dynamic MTs, with periods of slow growth, pausing, and transitions to rapid depolymerization (catastrophes). The mean results for catastrophe counts in more than 24 
microtubule-hours of timelapse images are presented in Table \ref{tab:cats}. Using a Poisson test in the R statistical programming package, I determined that the number of catastrophes for each condition can not be explained by a single catastrophe frequency (p =  5.76e-10). 

\begin{table}[!ht]
\centering
\caption{
{\bf Catastrophe rates for unlabeled and 6\% Rhodamine-labelled dynamic MTs}}
\begin{tabular}{|l|c|}
\hline
 {\bf Catastrophe rates} &  \\ \hline
{ Rhodamine-Labeled}  & 0.03662 min$^{-1}$  \\ \hline
{ Unlabeled} &  0.1477 min$^{-1}$  \\ \hline
{\bf Ratio} & 0.116 to 0.477 (99\% C.I.) \\ \hline
\end{tabular}
\label{tab:cats}
\end{table}

The 6\% rhodamine-tubulin labeling density I used in this experiment is well within the range of  concentrations reported in the literature \cite{hunt1998}. Label-free methods such as the one presented here offer a way to compare the effects of {\itshape in vitro} conditions (including recovery tags or fluorescence labels) for their effect on microtubule dynamics against MTs in as nearly native a form as possible. 

\subsubsection*{TIEM Achieves Greater SNR with Fewer Frames than VE-BF} 


I compared SNRs calculated for the same region of interest in corresponding video-enhanced bright field (VE-BF) and Transport of Intensity Equation (TIE) phase imaging. In fact, the VE-BF images are the inputs to the TIE algorithm with a separately acquired background subtracted. In TIE microscopy the background is effectively removed when taking the difference of the input images, so no separate background has to be acquired.
The SNRs for each imaging mode are presented in Table \ref{tab:SNRs}, where TIEM images of microtubules always have a greater SNR than an equivalent VE-BF image, and the frames required to generate a useful image is lower in TIEM. 
\begin{table}[!ht]
\centering
\caption{{\bf SNRs measured for VE-BF and TIEM images of MTs}}
\begin{tabular}{l |c |c | l  }
	{\bf Vertical MT } & {\bf SNR} & {\bf Frames } & \\ \hline
	VE-BF Defocused 225 nm & 1.0668 & 18 & \\
	$\phi(x,y)$ & 2.8910 & 18 & \\
	VE-BF Defocused 225 nm & 0.9360 & 28 & \\
	$\phi(x,y)$ & 3.7821 & 28 & \\
	VE-BF Defocused 225 nm & 0.9734 & 42 & \\
	$\phi(x,y)$ & 3.7431 & 42 & \\	
	{\bf Horizontal MT }&  &  & \\ \hline
	VE-BF Defocused 225 nm & 1.1724 & 18 & \\
	$\phi(x,y)$ & 3.2406 & 18 & \\
	VE-BF Defocused 225 nm & 1.3412 & 28 & \\
	$\phi(x,y)$ & 3.3045 & 28& \\
	VE-BF Defocused 225 nm & 1.7841 & 42 & \\
	$\phi(x,y)$ & 4.0716 & 42 & \\
	\label{table:SNR}
	\label{tab:SNRs}
	\end{tabular}
\label{table:SNR}
\end{table}

\begin{figure}[!h]
\includegraphics[keepaspectratio,scale=0.5]{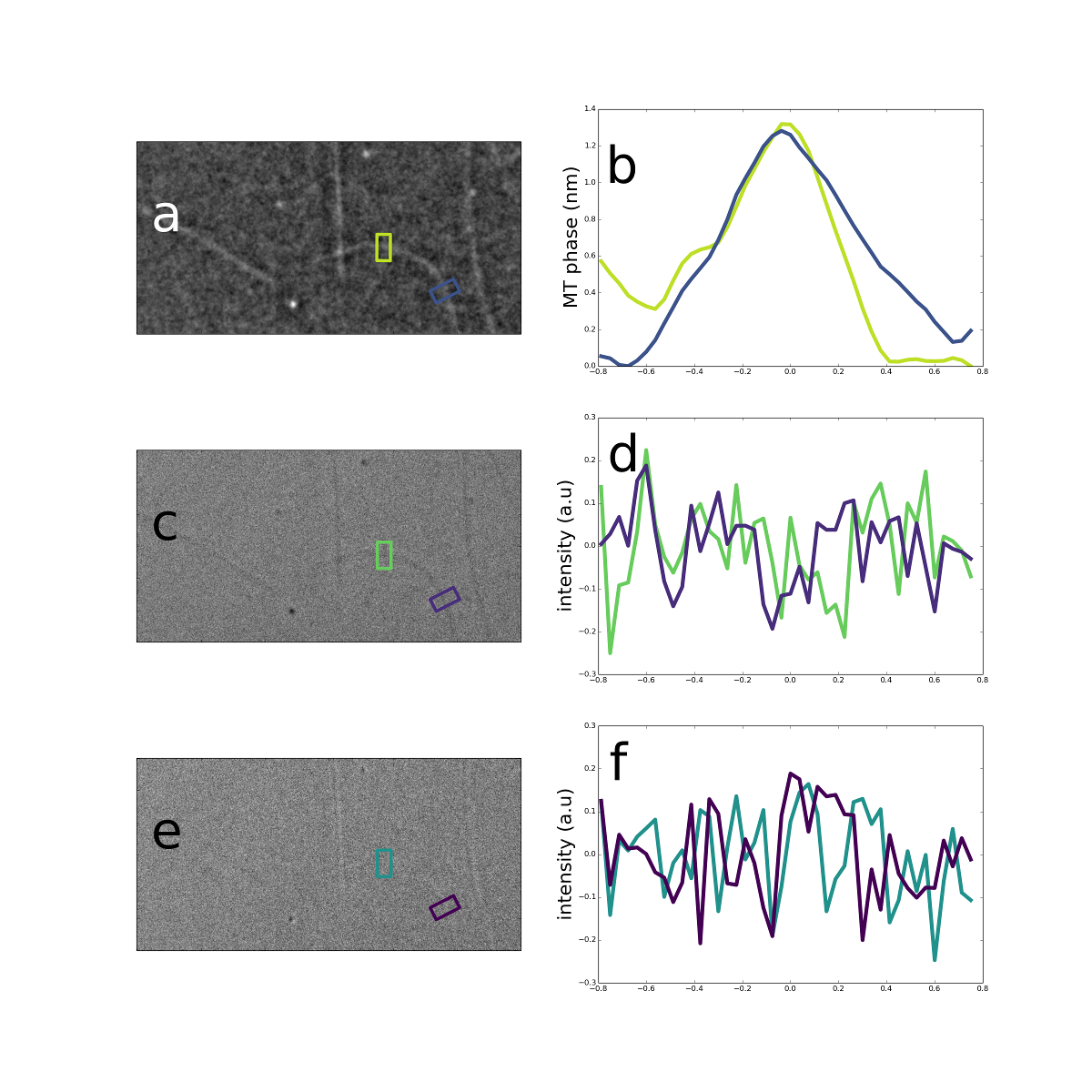}
\caption{{\bf TIE phase and VE-BF images used for calculating signal to noise ratios.}
{\bf (a)} TIE phase map of microtubules. 
Colored boxes indicate MT sections used to calculate the SNRs in Table 1. 
{\bf (b)} VE-BF image defocused 225 nm below the focal plane.{\bf (c)}  VE-BF image defocused 225 nm above the focal plane. {\bf (d)} Profile plot of averaged MT cross-sections in TIE phase map. {\bf (e)} Profile plot of averaged MT cross-sections from -225 nm defocused VE-BF image. {\bf (f)} Profile plot of averaged MT cross-sections from +225 nm defocused VE-BF image }
\label{fig:SNRCrossSections}
\end{figure}

SNR is not an ideal metric for assessing MT image quality. A recognizable image of an MT may arise as a correlated pattern in the characteristic rod shape even with a peak-to-peak amplitude at the same scale as the background noise. As we see in Fig \ref{fig:SNRCrossSections}, the SNRs calculated for VE-BF microscopy in this way are sometimes below 1.0, even where MTs are clearly recognizable. A better metric for MT imaging utility is how well the MTs can be recognized and localized by an experienced human operator or algorithm. Without an effective and objective way to test this directly, I leave the final judgment on image utility up to the reader. This can be assessed by comparing the imaging modes in Fig \ref{fig:vissummary} and additional examples in supplementary Fig S6. 


\newpage
\section*{Conclusions}

TIE phase imaging is an accessible label-free method to increase contrast for sub-resolution phase objects, with considerable improvement in SNR over video-enhanced bright field images of the same scene with fewer frames (Table \ref{table:SNR}). This SNR improvement over VE-BF is obtained while retaining a minimal optical setup, and in an optical tweezers system this means the optical trapping beam is unperturbed, {\itshape e.g.} by DIC prisms. Contrast is more isotropic than that of VE-DIC, and there are no prisms to periodically re-align. Both VE-BF and VE-DIC rely on background subtraction, and the background image (an average of many frames focused about 3 microns below the coverslip) has to be updated regularly. In TIE microscopy, background is removed as a result of subtracting two images at different focus depths, as both images contain the optical system and sample background. Consequently, TIEM has no imaging overhead associated with updating a  background mask. The result of subtracting two images, equivalent to the Laplacian of the sample phase profile, is qualitatively similar to Laplacian Phase Microscopy and can be used as a low-computation imaging mode for visualizing microtubules.

TIEM shares the benefits of competing and complementary label-free MT imaging methods, and offers the additional advantage of a minimal setup. In an optical tweezers system, TIEM avoids phototoxicity and photobleaching that would be inherent to extended live viewing with fluorescence microscopy. In an optical tweezers system, moving the sample to defocus the image does reduce the ability to record force measurements and TIE images simultaneously. However, this limitation could be circumvented by one of several motion-free options for defocusing, such as using a tuneable lens \cite{zuo2013}, a multifocus diffraction optical element \cite{abrahamsson2012} or a spatial light modulator \cite{maurer2010}.

My favorite option for defocusing without moving the sample, attractive for its simplicity, is the use of chromatic aberrations as defocused images. This was previously demonstrated for TIE microscopy of cells and test patterns \cite{waller2010}. I speculate that this could be implemented for TIEM of MTs based on single exposures using a white light source and RGB color camera as in \cite{waller2010}, or by sequentially switching the illumination wavelength.

The minimum hardware requirements for TIEM MT imaging are a standard brightfield microscope capable of focusing at increments of a few hundred nanometers. The optics for the enhanced phase imaging I demonstrate in this report are entirely computational; this technique can be widely adopted on existing microscopes already in use. 

\section*{Supporting Information}

Supporting information, including videos, python code, and interactive figures can be found at \url{https://github.com/qtzd/TIEMicroscopyOfMTs}

\paragraph*{{\bf S1: Video: Dynamic, Unlabeled MTs Visualized with TIEM. } }

\paragraph*{{\bf S2: Video: Dynamic MTs Labeled with Rhodamine and Visualized with TIEM. }}

\paragraph*{{\bf S3: Jupyter Notebook Interactive Figure S3: Exploring Offset Methods for Stabilizing  the 1D TIE in the Fourier Domain}}
{\bf (a)} The quadratic (found in the Fourier domain TIE) demonstrates the instability of the TIE. As the spatial frequency goes to 0, the FD TIE will blow up. We can add a small global offset to the denominator to prevent division by 0, or as we see here we can add a small sinusoidal bump (a shifted Hanning function) to selectively offset the quadratic for low spatial frequencies. 
{\bf (b)} The Fourier domain TIE demonstrates the effect of adding a Hanning bump (green) vs. a global offset (red). 
{\bf (c)} Simulated signal + noise used in this figure. 
{\bf (d)} Signal + noise recovered after convolution with {\bf b}.
{\bf (e)} Recovered signal with simulated noise removed.
To adjust width and amplitude of the Hanning bump, use the blue sliders at the bottom of the figure

\paragraph*{{\bf S4: Jupyter Notebook Interactive Figure : Effect of Parameter Choice on TIE Phase Micrograph of Rod Phantom}}
Fourier domain Poisson solvers and recovered TIE micrographs for a simulated rod phantom.
{\bf (a)} Fourier domain Poisson solver with a Hanning bump offset 
{\bf (b)} Fourier domain Poisson solver with a global offset 
{\bf (c)} Rod phantom phase map recovered with Hanning bump TIE solver
{\bf (d)} Rod phantom phase map recovered with global offset TIE solver
Blue sliders can be used to adjust $w_0$ and offset amplitude. Although the signal-to-noise ratio is reported here according to the definition for MT micrographs described in [1], the optimal choice of parameters depends on one's objectives. Setting amplitude and $w_0$ aggressively can be used for high contrast images with MTs that are more easily recognizable to a human user or pattern-recognition software. If accurate optical path lengths are desired for quantitative analysis lesser offsets may be desirable. Furthermore, high amplitude and low $w_0$ Hanning bumps create an integrated Fourier band-pass filter in the TIE solver and should be minimized for quantitative images of larger phase objects such as cells

\paragraph*{{\bf S5: Jupyter Notebook Interactive Figure : Effect of Parameter Choice on TIE Phase Micrograph of MTs}}
Fourier domain Poisson solvers and recovered TIE micrographs of microtubules.
{\bf (a)} Fourier domain Poisson solver with a Hanning bump offset 
{\bf (b)} Fourier domain Poisson solver with a global offset 
{\bf (c)} Microtubules micrograph phase map recovered with Hanning bump TIE solver
{\bf (d)} Microtubules micrograph phase map recovered with global offset TIE solver
The choice of parameters for the Hanning bump width and amplitude has effects the quality of the computed phase map. The left column displays the Fourier domain TIE solver filter and the resulting phase map, and the right column displays the Fourier domain solver using a global offset and the resulting phase map. This figure can be used to make decisions about when to use a global offset and when and what parameter settings to use with a Hanning bump.
To adjust width and amplitude of the Hanning bump, use the blue sliders at the bottom of the figure

\paragraph*{{\bf S6: Comparison of VE-BF, TIEM, and Additional Computational Imaging Modes for Visualizing MTs}}

S6 is an interactive version of Figure 1: TIEM of microtubules and comparison to other imaging modes. Three examples of MT images are available for comparison, or the user can replace these with their own image stacks for comparison.

{\bf (a)}) Video-enhanced bright field micrograph of microtubules adhered to a coverslip (average of 42 frames with background subtracted), defocused below the sample plane 
{\bf (b)})  VE-BF at the sample plane ($I_0$)
{\bf (c)})  VE-BF  225 nm above the sample plane ($I_a$). {\bf d)} TIE phase contrast micrograph of microtubules.
{\bf (d)}) TIE phase contrast micrograph of microtubules.
{\bf (e)}) The gradient of {\bf d)} mimicking DIC contrast with a y-axis shear.
{\bf (f)}) Gradient of {\bf d)} mimicking DIC contrast with a x-axis shear. Contrast in {\bf e)}) and {\bf f)} is essentially nonexistent for MTs parallel to the shear axis.
{\bf (g)}) The Laplacian ($\nabla^2[\phi(x,y)]$) of {\bf d)}. 
{\bf (h)})  The intensity difference image ($I_a-I_b$). For micrographs of sufficiently thin phase objects, the first term in Eq. \ref{eq:productrule} goes to 0 and {\bf (h)} and {\bf g)} are mathematically equivalent. 
{\bf (i)}) Through the eyepiece' (no background subtraction) view of the sample plane.

\paragraph*{\bf S7: TIEMMT.py} {{\bf Python Code: TIE microscopy for microtubules}} This algorithm estimates a solution to the transport of intensity equation and returns a phase micrograph. The algorithm assumes a sufficiently thin phase object, as described in Eq. \ref{eq:productrule}. This code is included as part of the Jupyter notebook that produces Figures S3-S5.

\section*{Acknowledgments}
I gratefully acknowledge the efforts of Valerie Bentivegna and Thomas Rabl for counting microtubule catastrophes in over 24 MT-hours of blinded timelapse images, and Sascha Reidt for helpful commentary concerning the mathematics. Discussions with Rainer Heintzmann on the TIE were very helpful. David McGloin and Tomoyuki Tanaka of the University of Dundee provided equipment, reagents, and space. Harinath Doudhi, Valerie Bentivegna, Tomoyuki Tanaka, and David McGloin provided general comments on the manuscript.

\section*{Funding}
This project has received funding from the European Union's Seventh Framework Programme for research, technological development and demonstration under grant agreement no. 608133.

\end{document}